\documentclass[a4paper,11pt]{article}
\usepackage{jheppub} 
\usepackage{lineno}
\usepackage{natbib}

\providecommand{\dif}{\mathrm{d}} 
\newcommand{\beq}{\begin{equation}}
\newcommand{\eeq}{\end{equation}}
\newcommand{\bea}{\begin{eqnarray}}
\newcommand{\eea}{\end{eqnarray}}

\title{Radiating particle in the vicinity of the weakly charged Schwarzschild black hole}

\author[a]{Bakhtinur Juraev,}
\author[a]{Zden\v{e}k Stuchlík,}
\author[a,b]{Arman Tursunov}
\author[a]{and Martin Kolo\v{s}}

\affiliation[a]{Research Centre for Theoretical Physics and Astrophysics, Institute of Physics, Silesian University in Opava, CZ-74601 Opava, Czech Republic}

\affiliation[b]{Max Planck Institute for Radio Astronomy, Auf dem H{\"u}gel 69, Bonn D-53121, Germany}

\emailAdd{bakhtinur.juraev@gmail.com}
\emailAdd{zdenek.stuchlik@physics.slu.cz}
\emailAdd{arman.tursunov@physics.slu.cz}
\emailAdd{martin.kolos@physics.slu.cz}

\abstract{It is well known that supermassive black holes in the centers of galaxies are capable of accelerating charged particles to very high energies. In many cases, the particle acceleration by black holes occurs electromagnetically through an electric field induced by the source. In such scenarios, the accelerated particles radiate electromagnetic waves, leading to the appearance of the backreaction force, which can considerably change the dynamics, especially, if the particles are relativistic. The effect of the radiation reaction force due to accelerating electric field of the central body in curved spacetime has not been considered previously. We study the dynamics of radiating charged particles in the field of the Schwarzschild black hole in the presence of an electric field associated with a small central charge of negligible gravitational influence. We start from the flat spacetime description, solving the Lorentz-Dirac equation reduced to the Landau-Lifshitz form. In curved spacetime, we use the DeWitt-Brehme equation and discuss the effect of the self-force, also known as the tail term, within the given approach. We also study the pure effect of the self-force to calculate the radiative deceleration of radially moving charged particles. In the case of bounded orbits, we find that the radiation reaction force can stabilize and circularize the orbits of oscillating charged particles by suppressing the oscillations or causing the particles to spiral down into the black hole depending on the sign of the electrostatic interaction. In all cases, we calculate the energy losses and exact trajectories of charged particles for different values and signs of electric charge.}

\begin{document}
\maketitle
\flushbottom



\section{Introduction} \label{sec-intro}

In recent years charged particle motion in the field of magnetized Schwarzschild or Kerr black holes was extensively studied, namely, in quasi-circular oscillatory regime or chaotic and acceleration regimes \citep{Kov-Kop-Stu:2010:CLAQG:, Kov-Stu-Kar:2008:CLAQG:, Fro-Sho:2010:PHYSR4:, Kol-Stu-Tur:2015:CLAQG:, Stu-Kol:2016:APJ:, Tur-Stu-Kol:2016:PRD:, Kol-Tur-Stu:2017:EPJ:, Pan-Kol-Stu:2019:EPJ:, Stu-Kol-Tur:2021:Universe:, Pug-Stu:2021:CLAQG:,Pug-Stu:2021:PASJ:}. Magnetic Penrose process \citep{Wag-Dhur-Dad:1985:APJ:,Zaj-Tur-Ekc-Bri:2018:MNRAS:,Stu-Kol-Tur:2021:Universe:} was demonstrated as a possible explanation of extremely energetic particles accelerated in ionized Keplerian disk orbiting supermassive black holes ($M=10^{9} M_{\odot}$) immersed in magnetic field of intensity $B \sim 10^{4} G$ \cite{Tur-Stu-Kol-Dad-Ahm:2020:APJ:}. It is significant that in such acceleration processes, the influence of the back-reaction due to synchrotron radiation of such particles can be neglected for protons (or ions) but can be strongly significant for electrons \cite{Stu-Kol-Tur:2021:Universe:}; moreover, due to radiative processes in the ergosphere of Kerr black holes, the so-called radiative Penrose Process, increasing energy of radiating counter rotating particles, can enter the play \cite{Kol-Tur-Stu:2021:PRD:}. 
However, surprisingly, it was demonstrated that an electric Penrose process related to a supermassive weakly charged black hole is able to explain such highly energetic particles too \cite{Tur-Jur-Stu-Kol:2021:PRD:}. It is thus of high interest to estimate the influence of the back-reaction due to radiating charged particles moving around electrically charged black holes.

The radiation reaction force of particles moving in the field of weakly charged black holes appears when a charged particle emits bremsstrahlung radiation. This paper aims to study test particle motion in combined electric and gravitational fields, taking radiation reaction force into account. Similar studies for the Schwarzschild black hole immersed in an external magnetic field were presented in \cite{Tur-Kol-Stu-Gal:2018:APJ:,Tur-Kol-Stu:2018:AN:,Shoom:2015:PRD:,2020Univ....6...26S}. 

The no-hair theorem of black hole physics states that black holes may be characterized by three parameters --  mass, spin, and charge. In many astrophysical scenarios, the charge of the black hole is neglected, on the one hand because of unrealistically large charge values required for sensible effect on the spacetime metric and, on the other hand, due to rapid discharge of any net-charge excess by accretion of a plasma surrounding the black hole. The maximal theoretical limit on the charge can be obtained from the comparison of the gravitational radius of a black hole with the characteristic length of the charge $Q_{G}$, which gives  
\begin{equation}
    \sqrt{\frac{Q^{2}_{G} G}{c^{4}}} = \frac{2 G M}{c^{2}} \Rightarrow Q_{G} = 2 G^{1/2} M \approx 10^{30} \frac{M}{M_{\odot}} \text{Fr.}
\end{equation}
In realistic scenarios, however, such a strong charge in unachievable. 
In esu-cgs system of units, the electrostatic unit of charge is given by
\begin{equation*}
   1 \, {\rm Fr} \equiv 1 \, {\rm esu} = 1 \, {\rm cm}^{3/2} {\rm g}^{1/2} {\rm s}^{-1}, \quad 1 \, {\rm C} = 3 \times 10^{9} \,\, {\rm Fr}.
\end{equation*}
On the other hand, there exist several astrophysical scenarios based on charge separation and selective accretion, in which a black hole can acquire a small electric charge \citep{Ruf-Wil:1975:PRD:, Bal-Har:1978:APJ:,Wei-Dra-Bar:2006:APJ:,Wald:1974:PHYSR4:,Zaj-Tur:2019:Observatory:, Zaj-Tur-Ekc-Bri:2018:MNRAS:,Lev-Dor-Gar:2018:APS:,Tur-Zaj-Eck-Bri-Stu-Cze:2020:APJ:,2023PhRvD.107d4055A}. For example, one of the charge limits can be obtained from a simple argument based on the mass difference between protons and electrons. The balance between gravitational and electrostatic forces for particles near the black hole is achieved when the black hole develops a positive net electric charge of the order $Q\sim3\times10^{11}~{\rm Fr}$ per solar mass \citep{Bal-Har:1978:APJ:,Zaj-Tur:2019:Observatory:}. Another charging mechanism is based on the irradiation of photons, ionizing and charging the matter surrounding a black hole by extracting electrons \citep{Wei-Dra-Bar:2006:APJ:,Stu-Kol:2016:EPJ:}. In such a situation, the black hole's charge is most certainly positive, of the order of $Q\sim10^{11}~{\rm Fr}$ per solar mass, which is of similar order of magnitude as in the above-mentioned estimate. According to the Wald's mechanism \cite{Wald:1974:PHYSR4:} -- the twisting of magnetic field lines caused by the frame-dragging effect of the black hole's rotation naturally induces the charge. Consequently, the stellar mass black hole ($M=10~M_{\odot}$)  and its surrounding magnetosphere may have an equal and opposite charge of $Q\sim10^{18}~{\rm Fr}$ per solar mass. The range of realistic values of the black hole's charge, which may vary depending on whether the black hole is spinning or not, thus takes the form
\begin{eqnarray}
 10^{11} \frac{M}{M_{\odot}} \, {\rm Fr} \lesssim Q_{\rm BH} \lesssim 10^{18} \frac{M}{M_{\odot}} \, {\rm Fr}. \label{BHchargelimits} 
\end{eqnarray}
The reader may also refer to \citep{Zaj-Tur-Ekc-Bri:2018:MNRAS:,Zaj-Tur:2019:Observatory:}, where the observational method for the black hole charge estimate has been proposed, based on the analysis of the bremsstrahlung surface brightness profiles in the case of the Galactic centre supermassive black hole.

The general properties of particle motion in the vicinity of electrically charged black hole were studied in \citep{Pug-Que-Ruf:2011:PRD:,Bal-Bic-Stu:1989:BAIC:,Bic-Stu-Bal:1989:BAIC:,Bis-Mer-Ruf-Ves:1993:APJ:}. The previous studies on the influence of the radiation reaction force on a charged particle moving around the black hole immersed in an external magnetic field \cite{Tur-Kol-Stu-Gal:2018:APJ:,Tur-Kol-Stu:2018:AN:,Shoom:2015:PRD:} showed that radiation reaction cannot be neglected in many physically interesting cases. We thus particularly expect their role also in the case of bremsstrahlung radiation of particles moving in the background of a weakly charged Schwarzschild black hole.

The structure of the paper is organized in the following order. In Sec.~\ref{RR_FLAT}, we derive the dynamical equations with the radiation reaction force of the charged particle in the Minkowski metric in spherical coordinates. In Sec. \ref{PARTICLE_MOTION_ELECTRIC_WR}, we study the charged particle dynamics without radiation reaction force in the vicinity of the weakly charged non-rotating Schwarzschild black hole. In Sec.~\ref{RADIATION_REACTION_SECTION}, we present a general relativistic treatment of the radiation-reaction force in curved spacetime. We derive the equations of motion for radiating particle in the vicinity of the weakly charged black hole. In Sec.~\ref{RADIAL_MOTION}, we study the purely radial motion of the radiating particle. We include electromagnetic self-force on an electric charge into the equation of motion, as obtained by Smith and Will \cite{Smi-Wil:1980:PRD:}. In Sec.~\ref{MODIFICATION_OF_OSCILLATIONS}, we construct the trajectories of the radiating and non-radiating particles. In Sec.~\ref{A_R}, we calculate relativistic values of the radiation reaction parameter $k$ and electric interaction parameter $\mathcal{Q}$, which are broadly used in the paper, switching to the Gaussian units and discuss the astrophysical relevance of the model. Finally, we summarize our obtained results in Sec.~\ref{SUMMARY}. 

We use the spacetime signature $(-,+,+,+)$ and the $G = 1 = c$ system of geometric units throughout the paper. We use the constants in esu-cgs system of units explicitly for expressions with astrophysical relevance. 
In the graphics, we use red color for radiating particle and black color for non-radiating particle. The range of Greek indices is from 0 to 3.

\section{Radiation reaction in flat spacetime \label{RR_FLAT}}

\subsection{General formalism}

\begin{figure*}
\centering
\includegraphics[width=0.3\hsize]{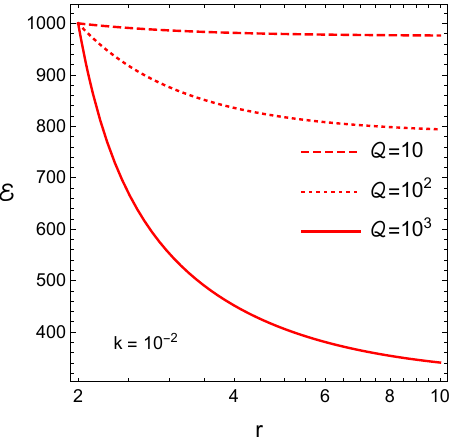}
\includegraphics[width=0.3\hsize]{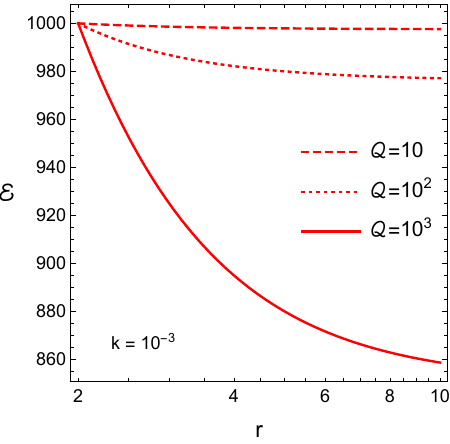}
\includegraphics[width=0.3\hsize]{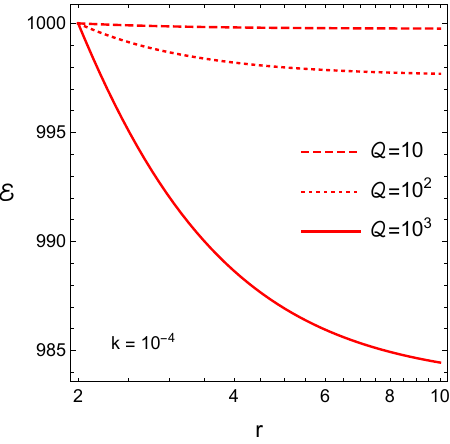}
\includegraphics[width=0.3\hsize]{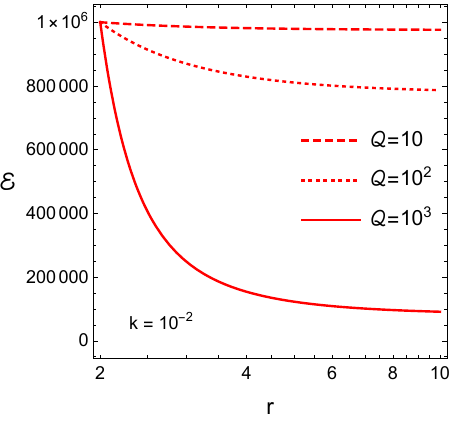}
\includegraphics[width=0.3\hsize]{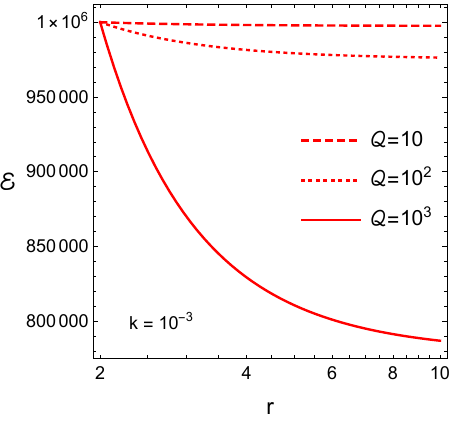}
\includegraphics[width=0.3\hsize]{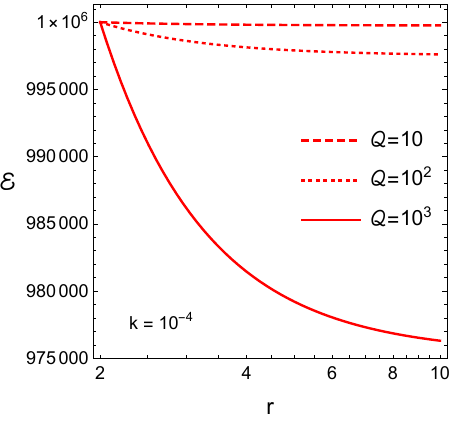}
\caption{Energy loss of a radiating charged particle in dependence on the distance $r$ from the central charge for different values of the electric interaction parameter $\mathcal{Q}$ and radiation parameter $k$. Here we compare 2 cases: The initial energy of the particle is $\mathcal{E}_{0} = 10^{3}$ (top row) and $\mathcal{E}_{0}=10^{6}$ (bottom row) of the rest mass of the accelerated particle. Note the different scales in $y$-axis of the presented plots. }
\label{Fig1}
\end{figure*}

For a more general understanding of the radiation reaction force caused by the electrostatic interaction, we first study the radiation reaction in a flat spacetime without gravity. The equation of motion of a charged particle in an electric field can be written as \cite{Tur-Kol-Stu-Gal:2018:APJ:}
\begin{equation}
    \frac{d u^{\mu}}{d \tau} = f^{\mu}_{L} + f^{\mu}_{R},
    \label{equation_of_motion_flat}
\end{equation}
where $f^{\mu}_{L} = (q/m)F^{\mu \nu} u_{\nu} $ is the Lorentz force and $f^{\mu}_{R}$ is the radiation reaction force. 
In the non-relativistic approximation the last term takes the form $\frac{2 q^{2} }{3m} \frac{d^{2} u^{\alpha}}{d \tau^{2}}$.  Using the condition $f^{\mu}_{R} u_{\mu} = 0$, one can obtain a special relativistic form of the equation for the radiation reaction force 
\begin{equation}
    f^{\mu}_{R} = \frac{2 q^{2}}{3m} \left( \frac{d^{2} u^{\mu}}{d \tau^{2}} +u^{\mu} u_{\nu} \frac{d^{2} u^{\nu}}{ d \tau^{2}} \right).
    \label{radiation_reaction_flat}
\end{equation}
The expression above is called the Lorentz-Abraham-Dirac equation or Lorentz-Dirac (LD) equation \cite{Dirac:1938:PRSLSA:}. The radiation recoil term, which corresponds to the relativistic correction of the radiation reaction force, is the second term in parentheses. The charged particle four-velocity satisfies the following conditions
\begin{equation}
    u_{\alpha} u^{\alpha} = -1, \quad u_{\alpha} \dot{u}^{\alpha} = 0, \quad u_{\alpha} \ddot{u}^{\alpha} = - \dot{u}_{\alpha} \dot{u}_{\alpha}.
\label{nor_con}
\end{equation}
The exact forms of the equations of motion (\ref{equation_of_motion_flat}) and (\ref{radiation_reaction_flat}) have disadvantages in practical calculations due to the exponential increase of the computational error. Reducing the order of the LD equation proposed in \cite{Landau:1975:CTP2:} can solve this problem. Alternatively, one can impose the Dirac’s asymptotic condition and integrate the higher-order LD equation backwards in time, as has been done in similar settings by \cite{Tur-Kol-Stu-Gal:2018:APJ:}. 
Substituting the derivatives of the Lorentz force for higher order terms in Eq.(\ref{radiation_reaction_flat}) yields the equation
\begin{equation}
    \frac{d u^{\mu}}{d \tau} = f^{\mu}_{L} + \frac{2 q^{2}}{3m} \left( \delta^{\mu}_{\alpha} + u^{\mu}u_{\alpha} \right) \frac{d f^{\alpha}_{L}}{d \tau}. \label{radiation_reaction_flat_2}
\end{equation}
This equation, known as the Landau-Lifshitz (LL) equation, has major implications: it is of the second order, does not contradict the principle of inertia, and the radiation reaction force disappears when the external (Coulomb) force is absent \citep{Roh:2001:PLA:,Poi:1999:Arx:}. In Eq.~(\ref{radiation_reaction_flat_2}), we can consider the Coulomb force as an external force acting on a charged particle and the equation for the radiation reaction in the case when $f^{\mu}_{L} = \frac{q}{m} F^{\mu \nu} u_{\nu}$ can be rewritten in the following form
\begin{equation}
    f^{\mu}_{R} = k \tilde{q} [F^{\mu \nu}_{,\alpha} + u^{\alpha}u_{\nu} +  \tilde{q} (F^{\mu \nu} F_{\nu \rho} - F^{\beta \alpha} F_{\beta \rho} u_{\alpha} u^{\mu})u^{\rho}],
\end{equation}
where $\tilde{q} = q/m$ is the specific charge of the particle and $k=(2/3) \tilde{q}q$ is the radiation reaction parameter. The comma in the first term inside the square brackets denotes the partial derivative with respect to the coordinate $x^{\alpha}$. More information on the problem of charged particle radiation reaction in flat spacetime may be found e.g. in \cite{Spo:2004:DOCPARF:}. 

\subsection{Radiation of charged particles in electric field}

In this section, we apply the above equations for a radiation of a charged particle in the field of a central fixed charge.  
The flat spacetime is given by the line element
\begin{equation}
    ds^{2} = - d t^{2} +  d r^{2} + r^2 d \theta^{2} + r^{2} \sin^{2} \theta d \phi^{2}.
\end{equation}
The only non-zero component of the electromagnetic four-potential $A_{\mu} = (A_{t},0,0,0)$ in the case of a central charge $Q$ has the following simple form
\beq
A_{t} = - \frac{Q}{r}. \label{aasbx}
\eeq
The antisymmetric tensor of the electromagnetic field $F_{\mu \nu} = A_{\nu,\mu} - A_{\mu,\nu}$ in this case has only two nonzero components
\beq \label{FaradayUniform}
F_{tr} = -F_{rt} = -\frac{Q}{r^{2}}.
\eeq
The difference between the LD and LL equations is in the number of initial conditions. In the case of the LD equation, nine constants must be set -- arbitrarily independent components of the charged particle initial position, velocity and acceleration. The other three constants are given by the normalization condition (\ref{nor_con}). Direct integration of higher-order equations causes the computational error to grow exponentially in a short period of time. Therefore, we apply the LL equation to govern the motion of radiating charged particles. It is worth noting that by integrating equations of motion backward in time, the problem of time computational error may be drastically decreased. There has been a proposal for a similar approach to solving Lorentz-Dirac equations in \citep{Bay-Hus:1976:PRD:, Tur-Kol-Stu-Gal:2018:APJ:}.

The non-vanishing components of the equations of motion (\ref{radiation_reaction_flat_2}) take explicitly the form
\bea
     \frac{d u^{t}}{d \tau} &=& \frac{\mathcal{Q} u^{r}}{r^{2}} - \frac{2 k \mathcal{Q} (u^{r})^{2}}{r^{3}},\label{EOFFLAT1}
     \\
   \frac{d u^{r}}{d \tau} &=& \frac{\mathcal{Q} u^{t}}{r^{2}} - \frac{2 k \mathcal{Q} u^{r} u^{t}}{r^{3}} , 
    \label{EOFFLAT2}
\eea
where $\mathcal{Q} = Qq/m$ is the electric interaction parameter and $k=2q^{2}/3m$ is the radiation parameter.

\subsection{Energy loss}
Radiated energy from the charged particle is the integral of radiation reaction force taken along the world line of the particle. The radiated energy can be found from the radiated four-momentum of the particle, which in flat spacetime has the following simple form $d P^{\mu}/d \tau = \frac{2}{3}q^{2} a^{\alpha} a_{\alpha} u^{\mu}$, or explicitly from Eq.(\ref{EOFFLAT1}). Using these equations, as demonstrated in \cite{Tur-Kol-Stu-Gal:2018:APJ:}, the energy loss of the radiated particle can be given as a function of $r$ by the relation 
\begin{equation}
    \frac{d \mathcal{E}}{d r} = - \frac{2 k \mathcal{Q}}{r^{3}} \sqrt{\left(\mathcal{E} - \frac{\mathcal{Q}}{r} \right)^{2} - 1} \, , 
    \label{loss_of_energy_flat}
\end{equation}
where $\mathcal{E}$ is the specific energy of the particle. We solve Eq.~(\ref{loss_of_energy_flat}) numerically and present the results corresponding to the energy loss of a charged particle accelerated in the field of a test charge in Fig.~\ref{Fig1} for different values of the electrostatic interaction parameter $\mathcal{Q}$ and the radiation parameter $k$.  

A general analytic solution of Eq.~(\ref{loss_of_energy_flat}) is not trivial. Therefore, for illustrative purposes, we present the solution of Eq.~(\ref{loss_of_energy_flat}) in the following two limits. 

As one of the limits, let us consider the situation where the initial energy of the particle is much higher than the electric interaction parameter divided by the distance, i.e. $\mathcal{E}\gg \mathcal{Q}/r$. This situation may occur in the case of ultra-relativistic particles accelerated by a relatively weak electric field. Then the solution of Eq.~(\ref{loss_of_energy_flat}) leads to the form
\begin{equation}
    \mathcal{E} = \mathcal{E}_{0} + \frac{2k \mathcal{Q}^{2}}{3} \left( \frac{1}{r^{3}} - \frac{1}{r^{3}_{0}} \right).
\end{equation}

Another way of obtaining an analytical solution is the inverse situation, where the electric interaction parameter is much greater than the initial energy of the particle. Then the solution of Eq.~(\ref{loss_of_energy_flat}) takes the following form
\begin{equation}
    \mathcal{E} = e^{k \mathcal{Q} \left( \frac{1}{r^{2}} - \frac{1}{r_{0}^{2}} \right)} \mathcal{E}_{0},
\end{equation}
where $\mathcal{E}_{0}$ and $r_{0}$ are initial energy and position of the charged particle. 

\section{Charged particle motion around weakly charged black hole \label{PARTICLE_MOTION_ELECTRIC_WR}}

\subsection{Equations of motion for non-radiating particle}

We study the dynamics of the non-radiating particle in the vicinity of a weakly charged Schwarzschild black hole, whose metrics in the spherical coordinates ($t,r,\theta,\phi$) is given by
\bea
 && d s^2 = -f(r) d t^2 + f^{-1}(r) d r^2 + r^2(d \theta^2 + \sin^2\theta d \phi^2), \label{SCHmetric} \\
%
%
		&& f(r) = 1 - \frac{2 M}{r}, \label{lapsefun}
\eea
where $M$ is the black hole mass.
Hereafter, without loss of generality, we put the mass of the black hole equal to unity, $M=1$. The only non-zero covariant component of the electromagnetic four-potential is given by Eq.~(\ref{aasbx}), matching with a flat spacetime case. The non-vanishing components of the electromagnetic tensor are $F_{tr} = - F_{rt} =  -Q/r^{2}$.

The motion of charged particles without the effect of the radiation reaction is described by the standard Lorentz equation in curved spacetime.

For the particle of mass $m$ and charge $q$ the Lorentz equation reads 
\beq \label{eqmonorad}
 \frac{D u^\mu}{d \tau} \equiv \frac{d u^\mu}{ d \tau} + \Gamma^\mu_{\alpha\beta} u^\alpha u^\beta = \frac{q}{m} F^{\mu}_{\,\,\, \nu} u^{\nu}, 
\eeq

where  $u^{\mu}$ is the four-velocity of the particle, normalized by the condition $u^{\mu} u_{\mu} = - 1$, $\tau$ is the proper time of the particle and components of $\Gamma^\mu_{\alpha\beta}$ are the Christoffel symbols.

The symmetry of the Schwarzschild black hole~(\ref{SCHmetric}) and the electric field give us a right to find the conserved quantities associated with the time and space components of the generalized four-momentum $P_\alpha = m u_\alpha + q A_\alpha$. Thus, the energy and axial angular momentum of the charged particle as assessed by an observer at rest at infinity are given by expressions representing  the constants of the motion
\bea
 \frac{P_{t}}{m} &=&- \mathcal{E} = - \frac{E}{m} = u_{t} - \frac{q Q}{m r} , \\
 \frac{P_{\phi}}{m} &=& \mathcal{L}=  \frac{L}{m} = u_{\phi}, \label{angmom}
\eea
where $\mathcal{E}$ and $\mathcal{L}$ denote specific energy and specific axial angular momentum of the charged particle. Due to the spherical symmetry of the background, 
we can fix the motion to the equatorial plane, $\theta = \pi/2$. 

The normalization condition $g_{\alpha \beta}u^{\alpha} u^{\beta} = -1$, leads to the following expression 
\begin{equation}
\left(\frac{d r}{d \tau} \right)^{2} = -f \left( 1 + \frac{\mathcal{L}^{4}}{r^{2} \sin^{2}{\theta}} \right) + \left( \mathcal{E} + \frac{\mathcal{Q}}{r} \right)^{2}.
\end{equation}
From the equation above, one can find the effective potential (we choose only a positive branch, see \cite{Bic-Stu-Bal:1989:BAIC:})
\begin{equation}
 \mathcal{E}^{+} = V_{eff} (r, \theta, \mathcal{L}, \mathcal{Q}) = \frac{\mathcal{Q}}{r} + \sqrt{f(r) \left( 1 + \frac{\mathcal{L}^{2}}{r^{2} \sin^{2}{\theta}} \right)}, \label{Veff:Electric charge}
\end{equation}
where $\mathcal{Q} = Qq/m$ is a parameter characterizing the electric interaction between the charges of the particle and the black hole.

The effective potential~(\ref{Veff:Electric charge}) shows clear symmetry $(\cal{L},\mathcal{Q})\leftrightarrow(-\cal{L},-\mathcal{Q})$ that allows to distinguish the following two situations 
\begin{itemize}
\item[+] {\it plus configuration} $\mathcal{Q}>0$ {\it repulsive configuration}. The charged particle is repulsed from the black hole by the Coulomb force.
\item[--] {\it minus configuration} $\mathcal{Q}<0$ {\it attractive configuration}
The charged particle is attracted to the black hole by the Coulomb force.
\end{itemize}
We choose the notations analogous to the weakly magnetized Schwarzschild black hole case \cite{Kol-Stu-Tur:2015:CLAQG:}.

\subsection{Radial and circular motion}
A special case of purely radial trajectories (motion) is governed by effective potential with vanishing angular momentum, $\mathcal{L}=0$, in Eq.(\ref{Veff:Electric charge}). Two special cases are interesting, namely the stationary points and circular orbits. 
\subsubsection{Stationary points} 
These correspond to the equilibrium position of charged particles, where the gravitational and electric forces balance each other. Clearly such points can exist outside the black hole for the repulsive electric force, case $\mathcal{Q}>0$, and they correspond to extrema of the effective potential $V_{\rm eff}(r,\mathcal{Q})$ for $\mathcal{L} = 0$, i.e., are given by $\partial_r V_{\rm eff}(r,\mathcal{Q},\mathcal{L}=0) = 0$. We thus arrive at the stationary points given by
\begin{equation}
    r_{\rm sp} = \frac{ 2 \mathcal{Q}^{2}}{Q^{2}-1} .
\end{equation}
The stability of the stationary points is determined by
\begin{equation}
  \partial_{rr} V_{\rm eff}(r, \mathcal{Q},\mathcal{L}=0) =  2 \mathcal{Q} + \frac{3-2r}{r f^{3/2}} <0  
\end{equation}
We thus see that all the stationary points are unstable (note that stable equilibrium positions are possible in more complex geometries \citep{Bic-Stu-Bal:1989:BAIC:,Stu:1980:BAIC:,Bla-Stu:2016:PRD:}). 
\subsubsection{Circular orbits} 
To find the circular orbits around the black hole, we have to solve simultaneously the following two conditions
\begin{equation}
    \partial_{r} V_{\rm eff}(r,\theta, \mathcal{L} ,  \mathcal{Q}) = 0 , \quad \partial_{\theta} V_{\rm eff}(r, \theta, \mathcal{L}, \mathcal{Q}) = 0. 
\label{stable_orbit_condtions}
\end{equation}
The motion in central fields, determined by Eq.~(\ref{Veff:Electric charge}), is always stable against perturbation in the $\theta$-direction, since $\partial_{\theta\theta}V_{\rm eff}>0$.

Due to the same spherical symmetry of the Schwarzschild black hole and the monopole electric field, one can fix the central plane of particle motion, for simplicity $\theta = \pi/2$.
The first extrema of Eq.~(\ref{stable_orbit_condtions}) gives us
\begin{equation}
    \mathcal{L}^2(r-3) + r^2 \left[ \mathcal{Q} \sqrt{f \left( 1 + \frac{\mathcal{L}^2}{r^2} \right)}-1 \right] = 0.
\end{equation}
The solution of the equation above gives the four roots of specific angular momentum $\mathcal{L}$
\begin{equation}
    \mathcal{L}^{2}_{\pm} = \frac{r^{2}f D^{2}_{\pm}}{2(r-3)^{2}} ,
\label{Specific_Angular_momentum_stable_orbit}
\end{equation}
where $D_{\pm}$ is 
\begin{equation}
    \quad D_{\pm} = \sqrt{ \mathcal{Q}^{2} + 2 r (r-3) \pm \mathcal{Q} \sqrt{\mathcal{Q}^{2} +4r(r-3)}}.
\label{D}
\end{equation}

The energy of the charged particle at the circular orbit is given by
\begin{equation}
    \mathcal{E}^{2}_{\pm} = 
    \frac{2 \mathcal{Q} (r-3) +\sqrt{2} r f  D_{\pm}  }{2r (r-3)},
\label{Energy_at_circular_orbit}
\end{equation}
where $D_{\pm}$ is given by equation (\ref{D}). Here $\pm$ indicates the maximum and minimum values of the particle energy at the circular orbit. The detailed studies of the particle motion in the vicinity of the weakly charged black hole were presented in \citep{Tur-Jur-Stu-Kol:2021:PRD:,Pug-Que:2011:PRD:}. We have to be careful in separating the positive-root and negative-root state \citep{Bic-Stu-Bal:1989:BAIC:,Bal-Bic-Stu:1989:BAIC:}.  

\section{Bremsstrahlung of charged particles in the vicinity of weakly charged black hole \label{RADIATION_REACTION_SECTION}}

\subsection{Radiation reaction force}

The motion of a relativistic charged particle is governed by the LD equation, which includes the influence of the external electromagnetic fields and corresponding radiation-reaction force. The last force arises from the radiative field of the charged particle, and the covariant equations of motion in general, can be written in the form
\beq \label{cureqmogen1}
   \frac{D u^\mu}{\dif \tau} = \tilde{q} F^{\mu}_{\,\,\,\nu} u^{\nu} + \tilde{q} {\cal{F}}^{\mu}_{\,\,\,\nu}  u^{\nu}, 
\eeq
where the first term on the right-hand side of Eq.~(\ref{cureqmogen1}) corresponds to the Lorentz force with electromagnetic tensor $F_{\mu\nu}$ given by (\ref{FaradayUniform}), while the second term is the self-force of the charged particle with the particle radiative field ${\cal F}_{\mu\nu} = {\cal A}_{\nu,\mu} - {\cal A}_{\mu,\nu}$. A term $D / d \tau = u^\mu D_\mu $ is the covariant proper time derivative. After several algebraic manipulations, one can find the explicit form of Eq.~(\ref{cureqmogen1}) for the motion of radiating charged particle in curved background in the form \citep{Poi:2004:LRR:,Tur-Kol-Stu-Gal:2018:APJ:,Hob:1968:AP:}
\bea 
&& \frac{D u^\mu}{\dif \tau} = \frac{q}{m} F^{\mu}_{\,\,\,\nu} u^{\nu} 
+ \frac{2 q^2}{3 m} \left( \frac{D^2 u^\mu}{d\tau} + u^\mu u_\nu \frac{D^2 u^\nu}{d\tau} \right) \nonumber \\ 
&& + \frac{q^2}{3 m} \left(R^{\mu}_{\,\,\,\lambda} u^{\lambda} + R^{\nu}_{\,\,\,\lambda} u_{\nu} u^{\lambda} u^{\mu} \right) + \frac{2 q^2}{m} ~f^{\mu \nu}_{\rm \, tail} \,\, u_\nu, 
\label{eqmoDWBH}  
\eea 
where the last term of Eq.~(\ref{eqmoDWBH}) is the tail integral given by Green function \citep{Poi:2004:LRR:,Gal:1982:JPMG:} 
\beq
f^{\mu \nu}_{\rm \, tail}  = \int_{-\infty}^{\tau}     
D^{[\mu} G^{\nu]}_{ + \lambda'} \bigl(\tau,\tau'\bigr)   
u^{\lambda'}(\tau') \, d\tau' .
\eeq
The integral in the tail term is evaluated over the past history of the charged particle, with primes indicating its prior positions. In this section, we neglect the last term in Eq.~(\ref{eqmoDWBH}), as justified for realistic scenarios by \cite{Tur-Kol-Stu-Gal:2018:APJ:}. 
However, we will recover the tail term and study its influence in Section \ref{sec:tail} for the case of radial motion, using the solution obtained by Smith and Will \cite{Smi-Wil:1980:PRD:}. Since we assume the Schwarzschild spacetime that is Ricci flat, the third term in Eq.(\ref{eqmoDWBH}) vanishes automatically. 

Thus, for the purpose of the present section, the equations of motion can be written in the following form
\beq \label{cureqmo1}
   \frac{\dif u^\mu}{\dif \tau} + \Gamma^\mu_{\alpha\beta} u^\alpha u^\beta = \tilde{q} F^{\mu}_{\,\,\,\nu} u^{\nu} + f_R^\mu,
\eeq
with the radiation reaction force given by
\beq f_R^\mu = \frac{2 q^2}{3 m} \left( \frac{D^2 u^\mu}{d\tau^2} + u^\mu u_\nu \frac{D^2 u^\nu}{d\tau^2} \right). \label{curfradrec}
\eeq
Introducing the four-acceleration as a covariant derivative of four-velocity, $a^{\mu} = D u^{\mu} / d\tau$, one can rewrite the term $D^2 u^{\mu} / d\tau^2$ as follows
\bea 
&&\frac{D^2 u^{\mu}}{d\tau^2} \equiv \frac{D a^\mu}{d \tau} = \frac{d a^\mu}{d\tau} + \Gamma^{\mu}_{\alpha\beta} u^\alpha a^\beta 
= \frac{d}{d\tau} \left(\frac{d u^\mu}{d\tau} + \Gamma^{\mu}_{\alpha\beta} u^\alpha u^\beta \right) + \Gamma^{\mu}_{\alpha\beta} u^\alpha \left(\frac{d u^\beta}{d\tau} + \Gamma^{\beta}_{\rho\sigma} u^\rho u^\sigma \right) 
\nonumber \\
&&= \frac{d^2 u^\mu}{d \tau^2} + \left( \frac{\partial \Gamma^{\mu}_{\alpha\beta}}{\partial x^\gamma} u^\gamma u^\beta + 3 \Gamma^{\mu}_{\alpha\beta}  \frac{d u^\beta}{d\tau} + \Gamma^{\mu}_{\alpha\beta} \Gamma^{\beta}_{\rho\sigma} u^\rho u^\sigma \right) u^\alpha. \nonumber \\
  \label{curdadtau}
\eea
Thus, the general relativistic radiation reaction force acting on the charged particles is described by Eqs.~(\ref{curfradrec}) and (\ref{curdadtau}). However, the full form of the equations of motion is highly non-linear, and, moreover, it leads to the existence of runaway solutions. One can avoid this problem similarly to the flat spacetime case, namely by reducing the order of differential equations. In the absence of the radiation-reaction force, the motion of the charged particle in the external electromagnetic field is governed by Eq.~(\ref{eqmonorad}). Taking the covariant derivative with respect to the proper time from both sides of Eq.~(\ref{eqmonorad}), we get
\beq \label{curdadtauFab}
\frac{D^2 u^{\alpha}}{d\tau^2} 
= \tilde{q} \frac{D F^{\alpha}_{\,\,\,\beta}}{d x^{\mu}} u^\beta u^\mu + \tilde{q}^2 F^{\alpha}_{\,\,\,\beta} 
F^{\beta}_{\,\,\,\mu} u^\mu, 
\eeq
where the covariant derivative from the second-rank tensor reads
\beq
\frac{D F^{\alpha}_{\,\,\,\beta}}{d x^{\mu}} = \frac{\partial F^{\alpha}_{\,\,\,\beta}}{\partial x^{\mu}} + \Gamma^{\alpha}_{\mu\nu} F^{\nu}_{\,\,\,\beta} - 
\Gamma^{\nu}_{\beta\mu} F^{\alpha}_{\,\,\,\nu}.
\eeq
Substituting Eq.~(\ref{curdadtauFab}) into Eq.~(\ref{curfradrec}), we get the radiation reaction force in the covariant Landau-Lifshitz form \cite{Landau:1975:CTP2:}
\beq \label{curradforce}
f_R^\alpha = k \tilde{q} \left(\frac{D F^{\alpha}_{\,\,\,\beta}}{d x^{\mu}} u^\beta u^\mu + \tilde{q} \left( F^{\alpha}_{\,\,\,\beta} 
F^{\beta}_{\,\,\,\mu} +  F_{\mu\nu} F^{\nu}_{\,\,\,\sigma} u^\sigma u^\alpha \right) u^\mu \right),
\eeq
Thus Eqs.~(\ref{cureqmo1}) and (\ref{curradforce}) give a set of general relativistic equations of motion of charged particles in combined gravitational and electromagnetic fields, where the particles undergo the radiation-reaction force.
\begin{figure*}
\begin{center}
\includegraphics[width=7cm]{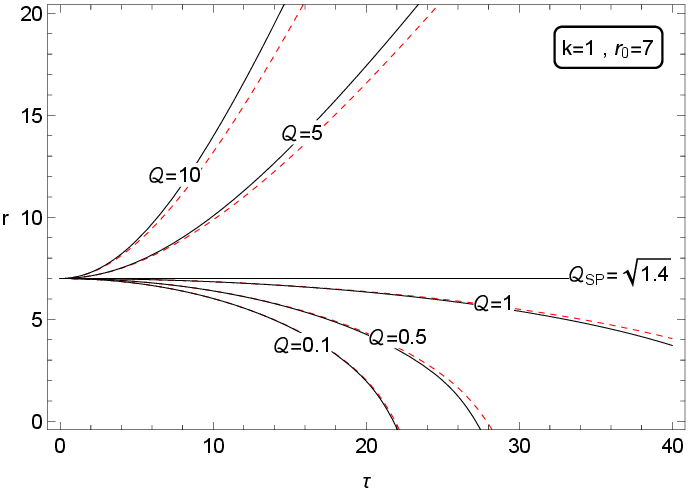}
\includegraphics[width=7cm]{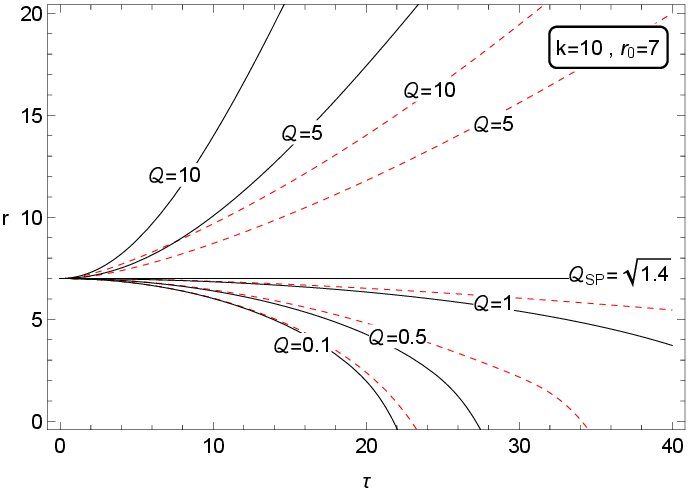}
\caption{Effect of the radiation reaction force on the charged particle radial $r$ position as a function of proper time $\tau$ for different values of radiation reaction parameter $k$ and electric interaction parameter $\mathcal{Q}$. The starting point of the particle is $r_{0}=7$. Equations of motion are given by Eqs.~(\ref{Radial_motion_eq_m_1}) and (\ref{Radial_motion_eq_m_2}). { Solid lines correspond to the motion without radiation reaction, while red dashed lines show the corresponding trajectories when radiation reaction is included. Radiation reaction force slows down the particle in both infalling and outgoing cases. The value $Q_{\rm SP}$ corresponds to the stationary case when the gravitational attraction is compensated by electrostatic repulsion due to $\mathcal{Q}>0$. The stationary cases are not influenced by the radiation reaction force, contrary to Fig.\ref{fig3}, where the tail term is taken into account. See discussion in Sec.\ref{RADIAL_MOTION}. }
\label{FIG2}} 
\end{center}
\end{figure*}

Fixing the plane of the motion at the equatorial plane $\theta=\pi/2$, we can find a simple explicit form of the equations of motion in our background. The non-vanishing components of equations of motion of radiating charged particle moving around a Schwarzschild black hole endowed with an electric monopole field take the following form

\begin{eqnarray}
    \frac{d u^t}{d \tau } &=&
\frac{\mathcal{Q} u^{r}}{f r^{2}} + \frac{\mathcal{Q} k}{f r^{3}} \left[ f r (u^{\phi})^{2} \{ r - \mathcal{Q} u^{t} \} - 2 (u^{r})^{2} \right]  -  \frac{2}{f r^{2}} u^{r} u^{t} ,
\label{EOF1} 
\\
\frac{d u^r}{d \tau } &=&
(u^{\phi})^{2} \{fr -1 \} - \frac{1}{r^{2}} - \frac{\mathcal{Q} k u^{r}}{r^{3}} \{  \mathcal{Q} r (u^{\phi})^{2} + 2 f u^{t} \}  + \frac{\mathcal{Q} f u^{t}}{r^{2}},
 \label{EOF2}
 \\
 \frac{d u^{\phi }}{d\tau } &=& - \frac{2 u^{\phi} u^{r}}{r} + \frac{\mathcal{Q} k}{r^{4}} u^{\phi} \{ f r u^{t} - \mathcal{Q} - \mathcal{Q} r^{2} (u^{\phi})^{2} \},
\label{EOF3} 
\end{eqnarray}
where $f$ is the Schwarzschild lapse function given by equation (\ref{lapsefun}), $\mathcal{Q} = Qq/m $ is the electric interaction parameter and $k=2q^{2}/3m$ is the radiation parameter.

\subsection{Energy and angular momentum loss}

The energy loss of the particle at the equatorial plane can be obtained from the equations (\ref{cureqmogen1}) and (\ref{curradforce}) leading to 
\begin{equation}
\frac{d \mathcal{E}}{d \tau} = - \frac{2 k \mathcal{Q}^{2}}{r^{3}} \mathcal{E}^{2} + \frac{k \mathcal{Q}^{2}}{r^{4}} \{4 - r^{2} (u^{\phi})^{2} \} \mathcal{E} - \eta ,
\label{Energy_loss}
\end{equation}
where $\eta$ is given by the equation
\begin{equation}
    \eta = \frac{k \mathcal{Q}}{r^{5}} [ 2 \mathcal{Q}^{2} - 2 f r^{2} - \mathcal{Q}^{2} r^{2} (u^{\phi})^{2} - 3 r^{4} f (u^{\phi})^{2}  ] .
\end{equation}
The first term in the Eq.~(\ref{Energy_loss}) becomes the leading term for the ultra-relativistic particle with $\mathcal{E} \gg 1$.

\begin{figure*}
\begin{center}
\includegraphics[width=7cm]{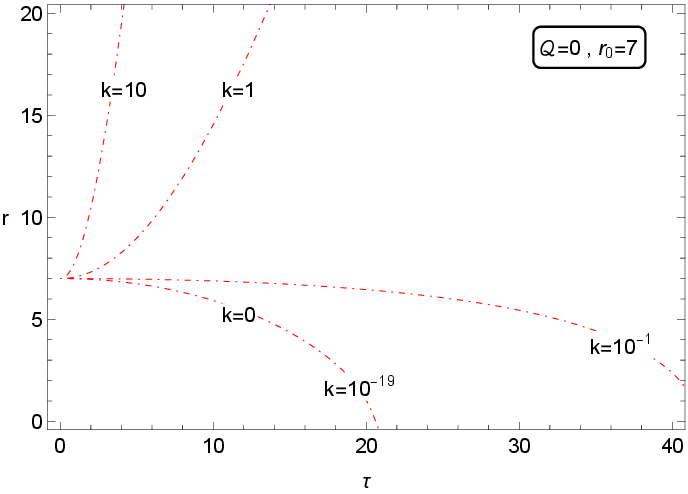}
\includegraphics[width=7cm]{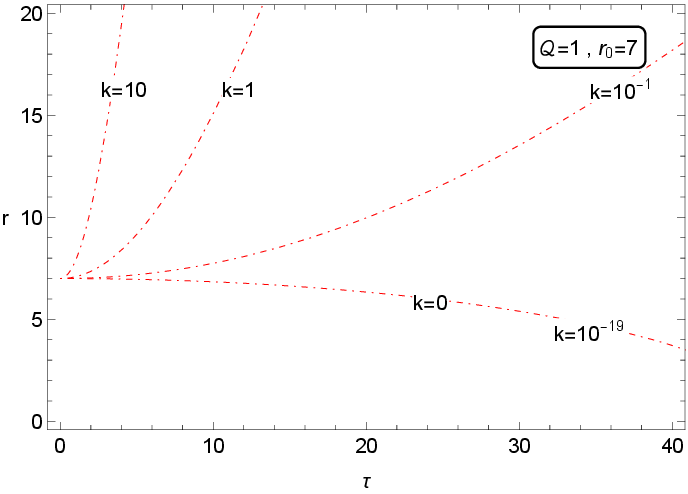}
\caption{Effect of the self-force on the radial motion of a charged particle without (left) and with (right) the presence of an electrostatic repulsion $\mathcal{Q}$, obtained by solving Eqs.~(\ref{Radial_motion_eq_m_3}) and (\ref{Radial_motion_eq_m_4}). 
Different values of radiation reaction parameter $k$  have been used. The starting point of the charged particle is $r_{0}=7$. See discussion in Sec.\ref{sec:tail}.
\label{fig3}} 
\end{center}
\end{figure*}

In a similar way, we obtain the equation for the loss of angular momentum of radiating charged particle 
\begin{equation}
\frac{d \mathcal{L}}{d \tau } = \frac{\mathcal{Q} k}{r^{2}} u^{\phi} \{ f r u^{t} - \mathcal{Q} - \mathcal{Q} r^{2} (u^{\phi})^{2} \}.
\end{equation}
In the next section, we study the influence of the radiation reaction on the trajectory of the charged particle moving around a weakly charged black hole.

\section{Radiative deceleration of the radially accelerated particles \label{RADIAL_MOTION}}

\subsection{Trajectories and stationary points} 

Let us consider a special case of the radial motion of radiating charged particle. Given the spherical symmetry of the spacetime and the central electric field, the radial motion scenario is physically well motivated. 
First, let us focus on the energy losses of the radiating particles in the radial motion case. The expression for the energy rate is reduced to the form
\begin{equation}
    \frac{d \mathcal{E}}{d \tau} =  -\frac{2 k \mathcal{Q}^{2}}{r^{3}} \mathcal{E}^{2} + \frac{4 k \mathcal{Q}^{2}}{r^{4}} \mathcal{E} - \frac{k \mathcal{Q}}{r^{5}} [2 \mathcal{Q}^{2} - 2 f r^{2}].
\end{equation}
The energy loss of the radiated particle as a function of a radial distance $r$ is thus given by the relation 
\begin{equation}
    \frac{d \mathcal{E}}{d r} = - \frac{ 2 k \mathcal{Q}}{r^{3}} \sqrt{\left( \mathcal{E}- \frac{\mathcal{Q}}{r} \right)^{2} - f}.
 \end{equation}
The angular momentum remains constant as its loss is given by $d \mathcal{L} / d \tau = 0$ in the radial motion in a weakly charged Schwarzschild metric.

Second, we focus on the treatment of the equations of motion. The non-vanishing components of equations of motion of radiating charged particle given by the Eqs.~(\ref{EOF1})-(\ref{EOF3}) can be rewritten for the purely radial motion in the following form 
\begin{eqnarray}
    \frac{d u^{t}}{d \tau} &=& \frac{\mathcal{Q}}{f r^{2}} - 2 \frac{\mathcal{Q} k}{f r^{3}} (u^{r})^{2} - \frac{2}{f r^{2}} u^{r} u^{t},
\label{Radial_motion_eq_m_1}
\\
    \frac{d u^{r}}{d \tau} &=& \frac{\mathcal{Q} f}{r^{2}} u^{t} - 2 \frac{\mathcal{Q} k f }{r^{3}} u^{t} u^{r} - \frac{1}{r^{2}}.
\label{Radial_motion_eq_m_2}
\end{eqnarray}
Solving the above equations numerically, we show in Fig.~\ref{FIG2} an influence of the radiation reaction on the motion of the particle comparing the trajectories with the non-radiating case. { We consider the case with $\mathcal{Q}>0$, which corresponds to the repulsive configuration, motivated by the particle acceleration models. 
Depending on the value of the interaction parameter $\mathcal{Q}$, the particle either falls into the black hole or escapes to infinity. There also exists a stationary case, where the gravitational attraction of the black hole is fully compensated by the electrostatic repulsion of the central charge. Radiation reaction leads to a slowing of the test particle in both infalling and outgoing cases. However, the stationary point is not influenced by the radiation reaction since, in this formalism, which neglects the tail term, a stationary particle does not radiate. }
In the next subsection, we will include electromagnetic self-force (tail term) in the equations of motion of the particle and test the influence of electromagnetic self-force on the trajectory of the particle in radial motion. 

\subsection{Effect of electromagnetic self-force} \label{sec:tail}

Equations of motion of a charged particle including the self-force in curved spacetime was introduced by DeWitt and Brehme in \cite{DeW-Bre:1960:AP:} and further corrected by Hobbs \cite{Hob:1968:AP:}. The solution of the equation for the electromagnetic self-force (tail term) acting radially on the electric charge without approximation for Schwarzschild black hole is derived by Smith and Will \citep{Smi-Wil:1980:PRD:,Poi-Pou-Veg:2011:LRR:}
\begin{equation}
    F^{r}_{\rm self} = \frac{q^{2}}{r^{3}} f^{1/2}, 
\label{EM_self-froce}
\end{equation}
where $f= 1-2/r$ is the lapse function. One should note that the electromagnetic self-force is repulsive \cite{Poi-Pou-Veg:2011:LRR:}. The components of the equations of motion of radiating particle for radial motion~(\ref{Radial_motion_eq_m_1})-(\ref{Radial_motion_eq_m_2})  taking into account the electromagnetic self-force given by Eq.~(\ref{EM_self-froce}) can be rewritten in the form
\begin{eqnarray}
    \frac{d u^{t}}{d \tau} &=& \frac{\mathcal{Q}}{f r^{2}} - 2 \frac{\mathcal{Q} k}{f r^{3}} (u^{r})^{2} - \frac{2}{f r^{2}} u^{r} u^{t},
    \label{Radial_motion_eq_m_3}
    \\
    \frac{d u^{r}}{d \tau} &=& \frac{\mathcal{Q} f}{r^{2}} u^{t} - 2 \frac{\mathcal{Q} k f }{r^{3}} u^{t} u^{r} - \frac{1}{r^{2}} + \frac{3 k f^{1/2}}{2 r^{3}}.
    \label{Radial_motion_eq_m_4}
\end{eqnarray}
The radial positions $r$ as a function of a proper time $\tau$ for the different values of radiation reaction parameter $k$ and for the absence and presence of electric parameter $\mathcal{Q}$ are plotted in Fig.~\ref{fig3}. One can notice that for the higher values of radiation reaction parameter $k$, electromagnetic self-force dominates gravitational force, which would cause an increase of the orbital radius $r$. It is caused due to the fact that electromagnetic self-force is repulsive. Similar results are obtained in \cite{Kom-Gor-Gar-Ver:2022:ARX:}. However, the radiation parameter $k$ is always very small in realistic situations, as we further discuss in Sec.~\ref{A_R}.

\section{Circular motion of the radiating charged particle\label{MODIFICATION_OF_OSCILLATIONS}}

\begin{figure*}
\begin{center}
\includegraphics[width=0.85\linewidth]{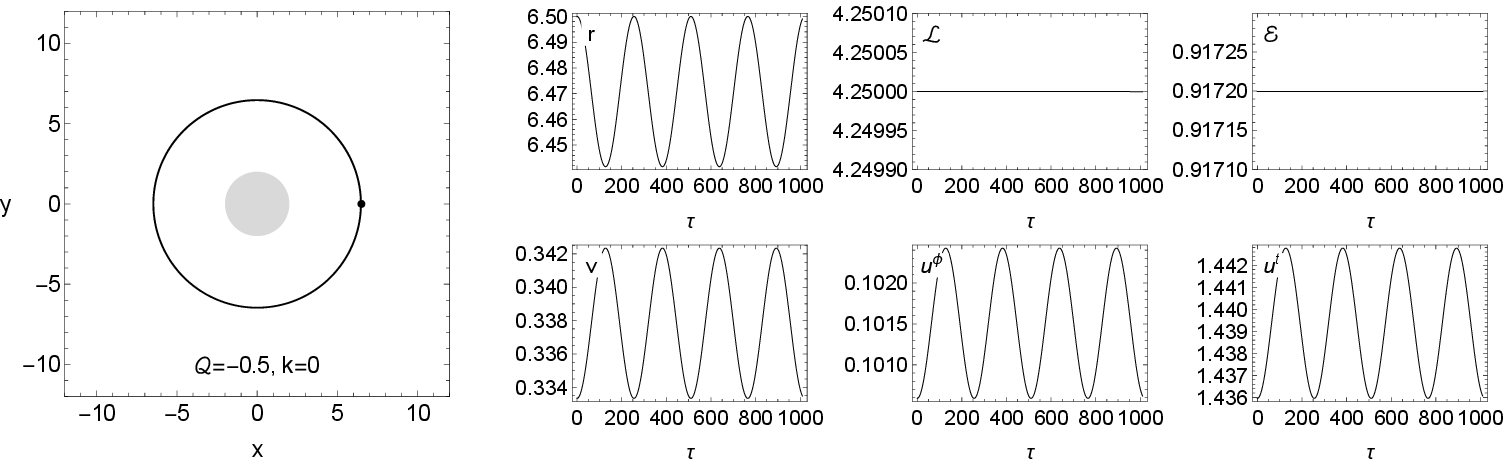}
\includegraphics[width=0.85\linewidth]{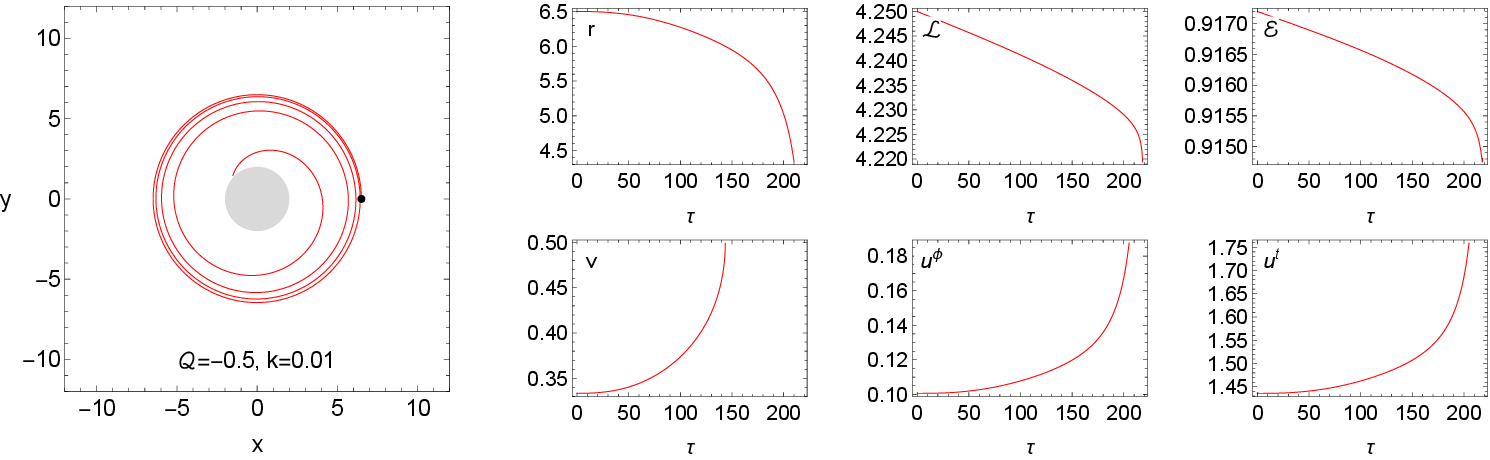}
\caption{Comparison of trajectories of non-radiating (black lines) and radiating (red lines) charged particles in the case of attractive Coulomb force around weakly charged black hole. Initial conditions for both cases are chosen the same. The black dot is the particle's starting point. The integration of the full set of equations of motion (\ref{cureqmo1}) -- (\ref{curradforce}) used for constructing the trajectories, fixing the plane of motion at $\theta = \pi/2 $. 
\label{fig-4}} 
\end{center}
\end{figure*}
\begin{figure*}
\begin{center}
\includegraphics[width=0.85\linewidth]{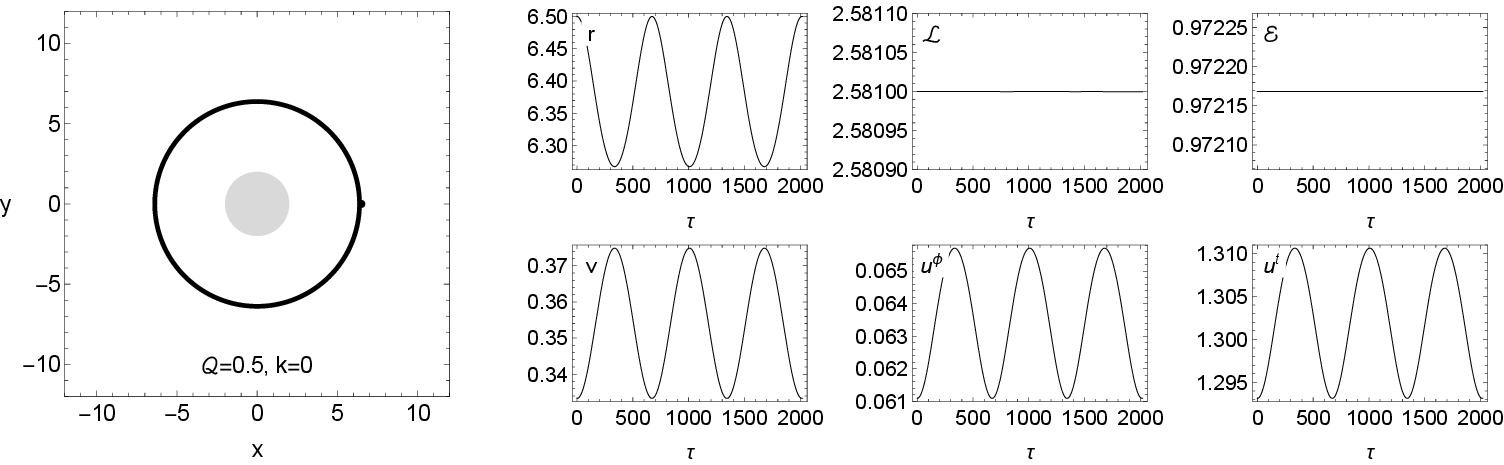}
\includegraphics[width=0.85\linewidth]{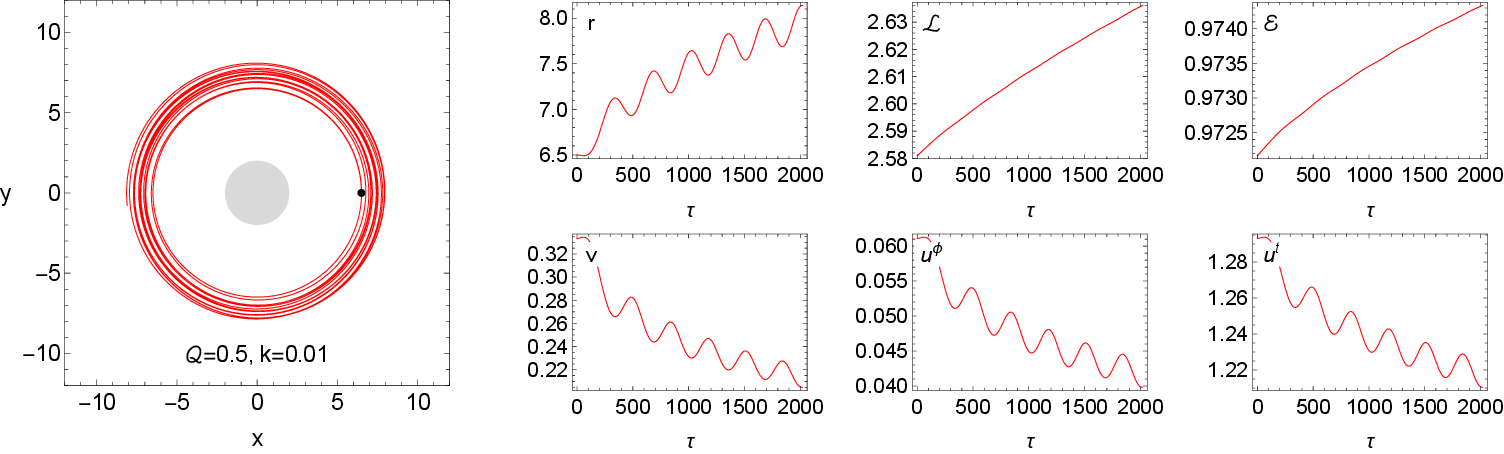}
\caption{Comparison of trajectories of non-radiating (black line) and radiating (red line) charged particles in the case of repulsive Coulomb force around a weakly charged black hole. The initial conditions for both cases are chosen the same way. Widening of the orbit can be observed. 
\label{fig-5}}
\end{center}
\end{figure*}

In this section we consider a more general, quasi-circular type of motion. 
The section aims to study the modifications of the charged particle epicyclic oscillations around circular orbits due to the radiation reaction force for "$+$" and "$-$" configurations. As mentioned in Sec.~\ref{PARTICLE_MOTION_ELECTRIC_WR}, "$+$" and "$-$" signs of the electric parameter $\mathcal{Q}$ indicates repulsive and attractive Coulomb force. 

Illustrative comparison of the trajectories of the non-radiating and radiating particles represented for the attractive Coulomb force in Fig.~\ref{fig-4} and for the repulsive force in Fig.~\ref{fig-5}. The trajectories in Fig.~\ref{fig-4} and Fig.~\ref{fig-5} are obtained by integrating the entire form of the Eqs.~(\ref{cureqmogen1})  and (\ref{curradforce}) fixing the plane of the motion at equatorial plane, as given in separated form in Eqs.~(\ref{EOF1})-(\ref{EOF3}).

The motion of the non-radiating particle is bounded for both cases. The oscillatory motion of the non-radiating particle caused by the initial energy and angular momentum of the particle corresponds to the state close but above the minimum of the effective potential. An impact of the radiation reaction force in the case of the attractive Coulomb force causes the loss of energy and angular momentum of the charged particle (see Fig.~\ref{fig-4}). Radiating particle accelerates due to the increase of the kinetic energy while losing its potential energy, which will cause the fall of the particle into the black hole. The situation takes a different scenario when the Coulomb force is repulsive(Fig.~\ref{fig-5}). The radiation reaction will cause the widening of the orbits of charged particle. This behavior of the radiating particle is explained by the deceleration of the particle, which will cause loss of its kinetic energy while increasing the value of potential one. Orbital widening for magnetized Schwarzschild black hole was studied in \cite{Tur-Kol-Stu:2018:AN:}. A detailed discussion of the phenomena is represented in the upcoming subsection.

\subsection{Orbital widening \label{Orbital_widening}}

An orbital widening effect of charged particles orbiting around a magnetized Schwarzschild black hole was studied in \citep{Tur-Kol-Stu-Gal:2018:APJ:,Tur-Kol-Stu:2018:AN:}. This effect describes the widening of orbits of the charged particles due to the influence of the Lorentz and radiation reaction forces. An analogous effect around electrically charged black holes exists only for repulsive Coulomb force. 
The radiated photons are emitted in the local geodesic frame of reference and are inside a small cone along the route of the charged particle for relativistic velocities, and hence, the radiation reaction force reduces the particle's kinetic energy and velocity. A decrease in the particle's angular velocity on a circular orbit can only be secured by an orbital radius expansion. As a result of this expansion, the particle's energy and angular momentum, as measured by an observer at infinity, grow as the particle is ascending in the effective potential. The specific energy of the particle reaches unity at asymptotic. 

The orbital widening effect for radiating charged particle around weakly charged Schwarzschild black hole is presented in Fig.~\ref{fig-5}. 

Recently, it has been argued in \cite{2023PhRvD.107f4046S} that the orbital widening effect reported for weakly magnetized black hole case in \cite{Tur-Kol-Stu-Gal:2018:APJ:,Tur-Kol-Stu:2018:AN:} is the result of a neglection of the self-force, i.e., the last tail term in the DeWitt and Brehme equation (\ref{eqmoDWBH}). However, due to the repulsive nature of the tail term, it seems unlikely that taking into account the tail term would switch the orbital widening off. Moreover, in the presence of the electric field, as studied in this paper, the tail term would add to the electrostatic repulsion. Therefore, in scenarios with the non-negligible tail term, the orbital widening effect would likely be enhanced in the full DeWitt-Brehme approach. 

\section{Relevance to astrophysics \label{A_R}}

Our studies of radiating charged particle motion are motivated by the various particle acceleration scenarios by black holes existing in the literature. In this section, we give estimates of the most relevant parameters of the described model applied to realistic situations.  

Restoring the world constants, we obtain dimensionless electric parameter $\mathcal{Q}$ in the form
\begin{equation}
    \mathcal{Q} = \frac{q Q}{G M m}.
\end{equation}
Possible values of the electric parameter compared to the radiation reaction parameter is given in Table~\ref{TabForce} for electrons, protons and partially ionized iron atoms. 

\begin{table}
\begin{center}
\begin{tabular}{| l | l  l  l  | l  l  l |}
\hline
 Q [Fr]  & & \quad $\mathcal{Q}$ & &  &  \quad $k \mathcal{Q}$ &  \\	 
\hline
 & \quad $e^{-}$ & \quad $p^{+}$  & \quad $Fe^{+}$ &  \quad $e^{-}$  & \quad $p^{+}$ & \quad $Fe^{+}$  \\ 
\hline
 $10^{30}$ & $\sim10^{20}$ &  $\sim10^{17}$ & $\sim10^{15}$ & $\sim10$ & $\sim10^{-5}$ & $\sim10^{-9}$ \\ 
 $10^{20}$ & $\sim10^{10}$ &  $\sim10^{7}$ &$\sim10^{5}$ & $\sim10^{-9} $ & $\sim10^{-15}$ & $\sim10^{-19}$
 \\
 $10^{15}$ & $\sim10^{5}$ &  $\sim10^{2}$ &$\sim10^{0}$ & $\sim10^{-14}$ & $\sim10^{-20}$ & $\sim10^{-24}$
 \\  
  $10^{11}$ & $\sim10$ & $\sim10^{-2}$ &$\sim10^{-4}$ & $\sim10^{-18}$ & $\sim10^{-24}$ & $\sim10^{-28}$
 \\  
  $10^{6}$ & $\sim10^{-4}$  & $\sim10^{-7}$ &$\sim10^{-9}$ & $\sim10^{-23}$ & $\sim10^{-29}$ & $\sim10^{-33}$
 \\  
 $10^{3}$ & $\sim10^{-7}$  & $\sim10^{-10}$ &$\sim10^{-12}$ & $\sim10^{-26}$ & $\sim10^{-32}$ & $\sim10^{-36}$
 \\  
\hline
\end{tabular}
\caption{Magnitudes of various forces acting on radiating charged particles in the dimensionless form of Eq.~(\ref{eqmoDWBH}) for various electric field strength values. The estimates are given for relativistic electron, proton and partially ionized iron atom (one electron lost) in the vicinity of stellar-mass black hole $M= 10 M_{\odot}.$ 
\label{TabForce}}
\end{center}
\end{table}

The dimensionless radiation reaction parameter $k$ already introduced in \cite{Tur-Kol-Stu-Gal:2018:APJ:}, take the form 
\begin{equation}
k = \frac{2 q^{2}}{ 3 m G M}.
\end{equation}
For electron, the radiation reaction parameter $k$ can take values 
\begin{eqnarray}
k_{\text{BH}} &\sim& 10^{-19} \quad \text{for} \quad M = 10 M_{\odot}, \\
k_{\text{SMBH}} &\sim& 10^{-27} \quad \text{for} \quad M = 10^{9} M_{\odot}, \\
k_{\text{Sgr}A^{*}} &\sim& 10^{-25} \quad \text{for} \quad M = 4.3 \times 10^{6} M_{\odot}.
\end{eqnarray}
The electric parameter $\mathcal{Q}$ and radiation reaction parameter $k$ for proton is lower by the factor $m_{p} / m_{e} \approx 1836$.

The equation of motion ~(\ref{eqmoDWBH}), describing the dynamics of a radiating particle, contains various terms of different magnitudes. The left-hand side of Eq.~(\ref{eqmoDWBH}) is the gravitational term, which is in the dimensionless units can be taken as $\sim 1$. So, all further terms can be considered as a relative ratio of the corresponding force to the gravitational force. The right-hand side of Eq.~(\ref{eqmoDWBH}) contains Coulomb force $\sim\mathcal{Q}$, radiation reaction force $\sim k \mathcal{Q}$ and electromagnetic self-force $\sim{k}$. Table~\ref{TabForce} shows the magnitudes of the Coulomb and radiation reaction forces in different physical scenarios. The range of possible values of electric charge includes the realistic estimates given by (\ref{BHchargelimits}). One can see that in some cases, the radiation reaction force can give a considerable contribution to the dynamical equations.

\section{Conclusions \label{SUMMARY}}

In this article, we have studied the radiation reaction of the particle in the vicinity of the non-rotating Schwarzschild black hole bearing a small electric charge, which does not affect the spacetime metric. The analysis started with deriving the equations of motion for charged particles in the flat spacetime, which is given by the Lorentz-Dirac and Landau-Lifshitz equations. The convenience of the Landau-Lifshitz equation gave us an opportunity to solve equations of motion in a more straightforward way. We have obtained equations of motion for charged particle in curved spacetime.
Further, we have used a solution derived by Smith and Will for electromagnetic self-force for the purely radial motion of the charged particle. 

To have a more intuitive description of the orientation of the Coulomb force directed inward and outward of the black hole, we have introduced plus and minus configurations, respectively. Our analysis showed that radiation reaction can influence the stability of the motion of the charged particle around the black hole.

We have studied the purely radial motion of the particle, including electromagnetic self-force in the equations of motion. We have shown that gravitational force dominates the electromagnetic self-force for the realistic values of the radiation reaction parameter $k$. On the other hand, due to the fact that electromagnetic self-force is repulsive, for higher values of radiation reaction parameter $k$, electromagnetic self-force takes on larger values than gravitational one, which causes an increase of the potential energy of the particle while losing kinetic one. 

We have shown that the appearance of the radiation reaction force can cause the widening of the circular orbit outwards of the black hole. This effect is possible only for repulsive Coulomb force. The appearance of the radiation reaction force along the circular orbit with repulsive electrostatic interaction decreases the linear velocity of the particle, which inevitably leads to the widening of the orbit. The widening of the orbit corresponds to the increase of the potential energy while the kinetic energy decreases.

In order to get a realistic view of the astrophysical relevance of the back-reaction phenomena related to the electromagnetic radiation of charged particle accelerated in the electric field of the black hole, we have provided the estimates of the relevant parameters of the theory, estimating electric interaction parameter $\mathcal{Q}$ and radiation reaction parameter $k$ for various values of the central charge and types of the test particle.

\section*{Acknowledgements}

This work is supported by the Research Centre for Theoretical Physics and Astrophysics, Institute of Physics, Silesian University in Opava, and Czech Science Foundation Grant No.~\mbox{23-07043S}. B.J. and Z.S. acknowledge the Silesian University in Opava Grant No. SGS/30/2023. A.T. acknowledges the Alexander von Humboldt Foundation for its Fellowship.


\bibliographystyle{JHEP}
\bibliography{biblio.bib}

\end{document}